\documentclass[usenatbib]{mnras}
\usepackage{graphicx}
\usepackage{amssymb}
\usepackage{amsmath}
\usepackage{journal_names}
\usepackage[usenames, dvipsnames]{color}

\title[SAMI outer dynamical profiles]{The SAMI Galaxy Survey: embedded discs and radial trends in outer dynamical support across the Hubble sequence}
\author[C. Foster et al.]{C. Foster,$^{1,2}$\thanks{E-mail: caroline.foster@sydney.edu.au} J. van de Sande,$^{1,2}$ L. Cortese,$^{2,3}$ S. M. Croom,$^{1,2,6}$ J. Bland-Hawthorn,$^{1}$\newauthor S. Brough,$^{4}$ J. J. Bryant,$^{1,2,5,6}$  M. Goodwin,$^{5}$ J. S. Lawrence,$^{5}$ N. Lorente,$^{5}$\newauthor A. M. Medling,$^{7,8,9}$ M. Owers,$^{5,10}$ S. N. Richards,$^{11}$ and N. Scott$^{1,6}$
\\
$^1$Sydney Institute for Astronomy, School of Physics, A28, The University of Sydney, NSW, 2006, Australia\\
$^2$ARC Centre of Excellence for All Sky Astrophysics in 3 Dimensions (ASTRO 3D)\\
$^3$International Centre for Radio Astronomy Research, University of Western Australia, 35 Stirling Highway, Crawley WA 6009, Australia\\
$^4$School of Physics, University of New South Wales, NSW 2052, Australia\\
$^5$Australian Astronomical Observatory, 105 Delhi Rd, North Ryde, NSW 2113, Australia\\
$^6$ARC Centre of Excellence for All-sky Astrophysics (CAASTRO)\\
$^7$Research School for Astronomy \& Astrophysics Australian National University Canberra, ACT 2611, Australia\\
$^8$Cahill Center for Astronomy and Astrophysics California Institute of Technology, MS 249-17 Pasadena, CA 91125, USA\\
$^9$Hubble Fellow\\
$^{10}$Department of Physics and Astronomy, Macquarie University, NSW 2109, Australia\\
$^{11}$SOFIA Operations Center, USRA, NASA Armstrong Flight Research Center, 2825 East Avenue P, Palmdale, CA 93550, USA\\
}

\begin{document}

\date{}

\pagerange{\pageref{firstpage}--\pageref{lastpage}} \pubyear{2018}

\maketitle

\label{firstpage}

\begin{abstract}

We study the balance in dynamical support of 384 galaxies with stellar kinematics out to $\ge1.5R_e$ in the Sydney AAO Multi-object Integral Field (SAMI) Galaxy Survey. We present radial dynamical profiles of the local rotation dominance parameter, $V/\sigma$, and local spin, $\lambda_{\rm loc}$. Although there is a broad range in amplitude, most kinematic profiles monotonically increase across the probed radial range. We do not find many galaxies with kinematic transitions such as those expected between the inner in-situ and outer accreted stars within the radial range probed. We compare the $V/\sigma$ gradient and maximum values to the visual morphologies of the galaxies to better understand the link between visual and kinematic morphologies.
We find that the radial distribution of dynamical support in galaxies is linked to their visual morphology. Late-type systems have higher rotational support at all radii and steeper $V/\sigma$ gradients compared to early-type galaxies.
We perform a search for embedded discs, which are rotationally supported discy structures embedded within large scale slowly or non-rotating structures. Visual inspection of the kinematics reveals at most three galaxies (out of 384) harbouring embedded discs. This is more than an order of magnitude fewer than the observed fraction in some local studies. Our tests suggest that this tension can be attributed to differences in the sample selection, spatial sampling and beam smearing due to seeing.

\end{abstract}


\section{Introduction}

Due to longer dynamical timescales, the outskirts of galaxies are thought to retain and reflect recent accretion events \citep[e.g.][]{Hoffman10,Wu14}. In contrast, deep potential wells efficiently funnel gas into the centres of galaxies enabling star formation. This has led to the expectation that the centres of early-type galaxies are primarily populated by stars formed in-situ, while the outskirts are dominated by accreted stars from past merger events. The observation of massive compact galaxies at high redshift \citep[e.g.][]{Daddi05,Trujillo06,vanDokkum08} combined with negative metallicity gradients and a tight age-[Mg/Fe] relationship at low redshift \citep[e.g.][]{MartinNavarro18} suggest that the bulk of the inner in-situ component formed early and rapidly through dissipational processes. The outskirts form over a more extended period in discs \citep[e.g.][]{Driver13} or are accreted through mergers with less massive companions \citep[e.g.][]{Oser12,Huang13}. The so-called ``two-phase galaxy formation'' scenario begins with an early and abrupt phase of intense in-situ star formation leading to a dense proto-galaxy at high redshift followed by a more progressive phase of predominantly minor merging which mainly build up the outer stellar component \citep[e.g.][]{Zhao03,Oser10, Oser12}.

A radial transition in the luminosity profiles and stellar kinematics between the inner in-situ and outer accreted stellar components of galaxies is thus expected \citep[e.g.][]{Cooper13,Foster16,Pulsoni17,RodriguezGomez17}. The transition radius may depend on mass \citep[e.g.][]{Cooper13,Pulsoni17}. Although other processes, such as stellar migration could blur the transition zone \citep[e.g.][]{ElBadry16}. Traditional integral field unit (IFU) studies have typically focused on the inner effective radius ($R_e$) of galaxies \citep[e.g.][]{Emsellem07,Cappellari11} and been unable to probe this transition. Notable progress has been achieved using the wide-field Mitchell Spectrograph \citep{Raskutti14,Boardman17} and the use of wide-field multiple slit spectrographs as ``proxy'' IFUs \citep[e.g.][]{Proctor09,Arnold11}. These studies have pushed 2D stellar kinematics in nearby galaxies out to $\sim 3R_e$, but sample sizes have been limited because they are observationally expensive.

Using sparsely sampled stellar kinematic data in the outskirts of nearby galaxies in the SAGES Legacy Unifying Globulars and GalaxieS (SLUGGS) Survey \citep{Brodie14}, \citet{Proctor09} showed that although NGC~821 and NGC~2768 have comparable spin (the so-called $\lambda$ parameter, e.g. \citealt{Emsellem07}) at one effective radius ($R_e$), the two differ markedly beyond $1R_e$. 
Using planetary nebulae as kinematic tracers, \citet{Coccato09} and \citet{Pulsoni17} also identified massive early-type galaxies with a marked decrease in rotational support beyond $1R_e$. \citet{Arnold11} found a significant fraction of elliptical galaxies with decreasing rotational support with radius. They interpreted this feature as an ``embedded disc'', i.e. a rotationally supported disc-like structure embedded within a larger scale slowly or non-rotating structure. Using the SAURON spectrograph, \citet{Weijmans09} confirmed the decreasing rotational support of the lenticular galaxy (S0) NGC 821 at large radii. However, \citet{Raskutti14} and \citet{Boardman17} did not find significant kinematic transitions in a sample of galaxies observed with the Mitchell Spectrograph. This discrepancy was attributed to differences in the sample selection of the respective studies.

Embedded discs and the bespoke two-phase kinematic transition have sometimes been used interchangeably to describe the same observed phenomenon \citep[e.g.][]{Arnold14,Bellstedt17a}. One can think of various formation pathways for embedded discs: e.g. major mergers \citep{Querejeta15b}, gas dissipation/accretion \citep[e.g.][]{White78} or the two-phase galaxy formation \citep{Oser12}. On the one hand, embedded discs are only identifiable by their radial decline in rotation, while on the other hand the two-phase galaxy formation scenario could accomodate a range of radial kinematic transitions given the diversity of possible merger orbital configurations and histories \citep[e.g.][]{Naab14} allowed. As such, embedded discs may be special cases of two-phase kinematic transitions.

In \citet{Bellstedt17a}, most embedded discs were found in either elliptical (E) galaxies or galaxies with uncertain morphological types, suggesting that the visual morphological classification of galaxies with embedded discs might be more representative of their \emph{global} rather than central (within $1R_e$) dynamical structure. Past studies have confirmed the link between the photometric properties of galaxies and their kinematic properties \citep[e.g.][]{Krajnovic13,Cortese16}. However, fast and slow rotator kinematic morphologies based on the  kinematics within $1R_e$ don't necessarily correspond to the lenticular and elliptical visual morphological types \citep[e.g.][]{Emsellem07,Emsellem11,Cappellari16a}.

While fast and slow rotator kinematic morphologies are based on a single kinematic measure, other studies have suggested that higher order kinematic moments are also a useful probe of a galaxy's dynamical state and better reflect their formation \citep{Naab14, vandeSande17a}. Similarly, \citet{Kalinova17} compared the shape of radial kinematic profiles with visual morphologies. In every case, some correspondence between visual and kinematic morphology was found, but all concluded that visual morphologies do not map directly onto kinematic morphologies.

\citet[][their fig. 11]{Arnold11} reported a trend between the global morphology of galaxies and the local spin gradient using SLUGGS data. They found that lenticular galaxies had steeper gradients than ellipticals. However, \citet{Raskutti14} did not find a relationship between morphology and the spin gradient in a sample of 33 massive elliptical galaxies observed with the Mitchell Spectrograph, even when combined with the SLUGGS data.  A reanalysis using kinemetry by \citet{Foster16} subsequently and independently confirmed the trend for the SLUGGS sample.

\begin{figure*}
\begin{center}
\includegraphics[width=150mm]{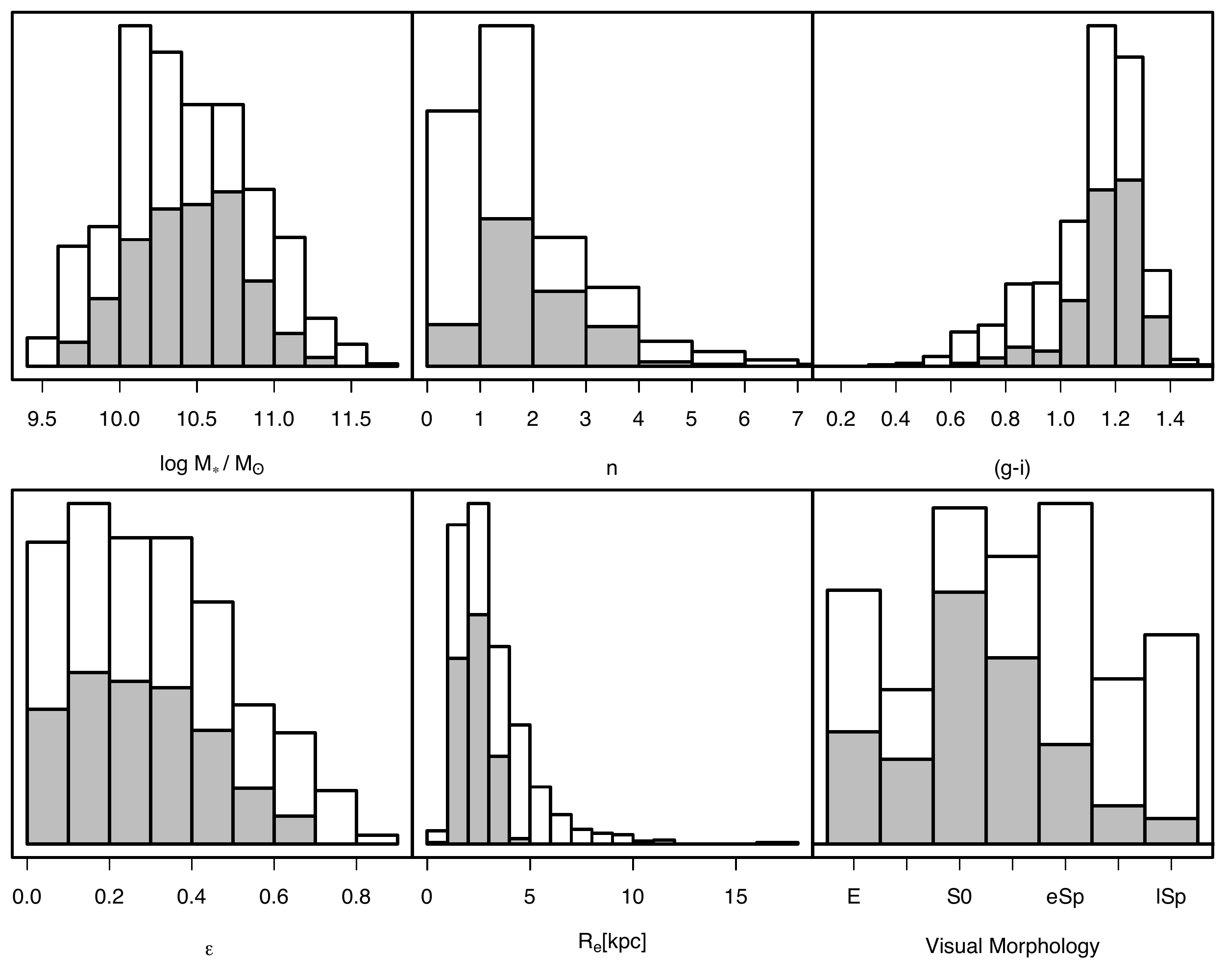}
\caption{Histograms showing the distributions in stellar mass $\log M_* / M_\odot$, S\'ersic index $n$, $(g-i)$ colour, apparent ellipticity $\epsilon$, effective radius $R_e$ and visual morphology (E -- elliptical, S0 -- lenticular, eSp -- early spiral, lSp -- late spiral) of the parent (white) and large scale kinematics (grey) samples. The large scale kinematics sample has no galaxies with effective radius larger than $\sim$5 kpc, a higher fraction of galaxies with high S\'ersic index and earlier morphological types than the parent sample.}\label{fig:hist}
\end{center}
\end{figure*}

In this work, we define a sample of 384 galaxies from the Sydney AAO Multi-object Integral Field (SAMI) Galaxy Survey with available large-scale stellar kinematics. We study the balance of dynamical support between rotation and random/pressure support. Although recent works have focused on the local \citep[e.g.][]{Wu14,Bellstedt17a} or central (within $R_e$) spin parameter $\lambda$ \citep[][]{Cappellari07,Emsellem07} over the classic and more intuitive $V/\sigma$ measurement \citep[e.g.][]{Binney78}, our work mainly shows the latter as both lead to qualitatively identical conclusions. Our emphasis is on large-scale $V/\sigma$ radial profiles, their slope and relation to morphological types. Special attention is given to contentious issues from the literature based on smaller samples already outlined. This study uses a wider morphological baseline (ellipticals to late-type spirals) and appreciably larger sample than past studies of large scale stellar kinematics, although we note that the visual morphological classifications \citep{Cortese16} are somewhat less certain owing to lower image quality and higher distances.

We assume a $\Lambda$CDM cosmology with  $\Omega_{\rm m}=0.3$, $\Omega_{\lambda}$~$=$~$0.7$ and $H_0=70$ km s$^{-1}$ Mpc$^{-1}$.

\section{Data}\label{sec:data}
The spectroscopic data used in this work were obtained using the SAMI instrument \citep{Croom12} as part of the SAMI Galaxy Survey. The SAMI instrument is a multiple integral field spectrograph capable of simultaneously obtaining spatially resolved spectroscopy of multiple targets using hexabundles \citep{BlandHawthorn11,Bryant14}. Each IFU hexabundle of 61 fibres has a 73\% filling factor with a 15 arcsec diameter. The SAMI instrument has 13 such hexabundles and 26 individual sky fibres over a 1 degree field-of-view. These are connected to the AAOmega spectrograph \citep{Sharp06} on the 3.9m Anglo-Australian Telescope. The observational campaign is now complete and the final SAMI Galaxy Survey contains $\sim3000$ low redshift ($0.004\le z\le0.095$) galaxies. The early, first and second public data releases are described in \citet{Allen15}, \citet{Green18} and Scott et al. (in preparation), respectively. Details of the SAMI target selection for the Galaxy And Mass Assembly (GAMA, \citealt{Driver11}) fields and cluster sample can be found in \citet{Bryant15} and \citet{Owers17}, respectively. 

The AAOmega spectrograph is set up with the 580V and 1000R gratings. This setup yields a wavelength coverage in the blue and red arms of 3750-5750 \AA\ and 6300-7400 \AA\ with a median spectral resolution of $R\sim1809$ and $R\sim4310$, respectively. The SAMI data are reduced by the {\sc 2dFdr} pipeline. The pipeline subtracts the bias frames and applies flat fielding, cosmic ray removal, wavelength calibration using CuAr arc frames and sky subtraction. Following the technique of \citet{Sharp10}, a spectrum is extracted for each fibre. For detail of the spectral data reduction steps in {\sc 2dFdr}, see \citet{Hopkins13}. Spectra are then flux calibrated using a primary standard star observed on the same night as the observations. Flux scaling and telluric absorption corrections are performed using a secondary standard star observed simultaneously with each field. To deal with gaps between fibres within the hexabundles and ensure continuous sampling, each field is typically observed with 7 dithers. For each galaxy, datacubes are reconstructed using a minimum of 6 dithers. Flux and covariance are carefully propagated onto a grid as described in \citet{Sharp15}. For full details of the SAMI data reduction, see \citet{Allen15} and \citet{Sharp15}.

The line-of-sight velocity distribution (LOSVD) for each spaxel is parametrised with a Gaussian using the penalised pixel-fitting ({\sc pPXF}, \citealt{Cappellari04, Cappellari16b}) algorithm. The velocity moments (i.e. the recession velocity, $V$, and velocity dispersion, $\sigma$) are measured by fitting template spectra for each spaxel. As described in \citet{vandeSande17a}, spectra within elliptical annuli are first combined into a high quality spectrum before fitting in order to determine the best template to subsequently use for fitting the LOSVD on corresponding member spaxels. This two-step fitting process mitigates uncertainties associated with possible template mismatch issues, especially in the lower signal-to-noise spectra. For each spaxel, the LOSVD ($V$, $\sigma$) is estimated through minimising the residuals between the observed spectrum and the corresponding pre-determined template broadened through convolution with a Gaussian.

We use the Multi-Gaussian Expansion technique \citep[MGE,][]{Emsellem94} and the code from \citet{Cappellari02} to measure the light-weighted global photometric ellipticity, position angle and effective radius ($R_e$) of each galaxy.  The MGE analysis is performed on Sloan Digital Sky Survey (SDSS) or VLT Survey Telescope (VST) 400 arcsec $r$-band cutout images. We first process each image with {\sc SExtractor} and {\sc PSFEx} \citep{Bertin11} to mask neighboring objects and build a model of the point spread function. We use the regularised version of MGE, which allows the fit to reflect the underlying light distribution rather than asymmetries such as bars, spiral arms or other non-axisymmetric features \citep{Scott13}. More detail on the MGE fitting can be found in D'Eugenio et al. (in prep).


Stellar masses are estimated from the $i$-band absolute magnitude and $(g-i)$ colours following \citet[][equation 8]{Taylor11}. Whenever available (i.e. for 98 percent of the galaxies in our sample), visual morphologies are taken from \citet{Cortese16}. Two percent of galaxies either have uncertain morphological types or have not yet been visually classified.

Our sample selection is as follows. Galaxies from the internal team release version v0.9.1 with no spaxels satisfying $V_{\rm error}<30$ km s$^{-1}$ and $\sigma_{\rm error} < 0.1\sigma + 25$ km s$^{-1}$ are discarded. From the remaining parent sample of 845 galaxies with reliable stellar kinematics, we identify a sample of galaxies with measured stellar kinematics ($V$ and $\sigma$) out to at least 1.5$R_e$. The outermost radius probed must have a minimum filling factor of 85 percent. There are 408 SAMI galaxies satisfying this selection. A further 24 galaxies are removed from the sample due to short profiles (see Section \ref{sec:analysis} for more detail). This yields a final sample of 384 galaxies. Fig. \ref{fig:hist} shows how the final large scale stellar kinematics sample compares with the parent sample. The selection is biased in favour of earlier morphological types and higher S\'ersic index values because stellar kinematics are more readily measured in systems where the stellar light dominates the flux budget (rather than emission from ionised gas). By construction, galaxies with effective radii larger than two thirds that of the field of view of an individual hexabundle will not make it into our sample. Fig. \ref{fig:hist} suggests that this angular size limit has prevented galaxies with $R_e\gtrsim5$ kpc from entering the final sample.

\section{Analysis and results}\label{sec:analysis}

\begin{figure*}
\begin{center}
\includegraphics[width=180mm]{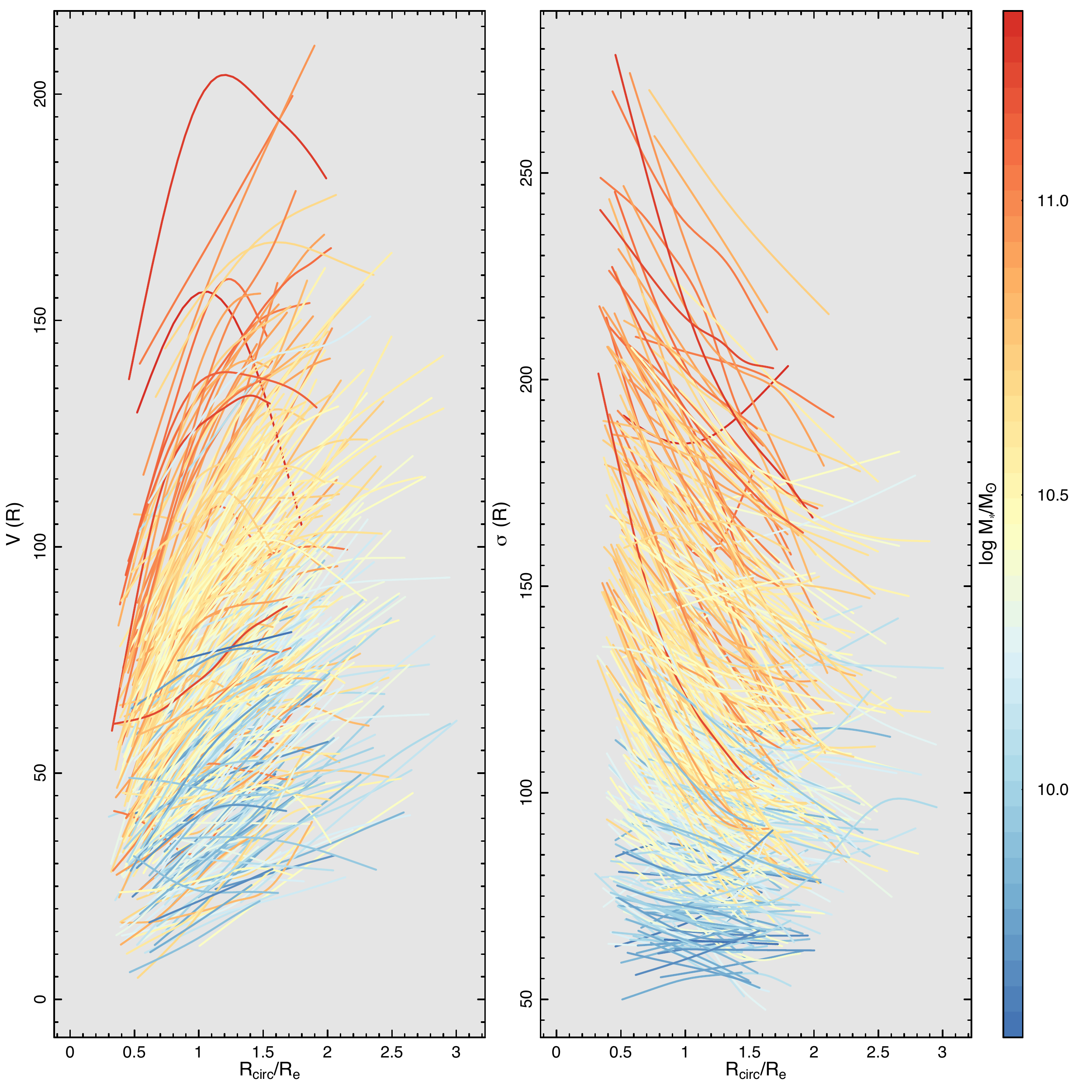}
\caption{Smoothed radial velocity $V$ (left) and dispersion $\sigma$ (right) profiles with measured stellar kinematics beyond $1.5R_e$. Profiles are colour-coded according to the galaxy stellar mass ($\log M_{*}$) as indicated by the colour scale on the far right. As expected, higher mass galaxies are able to sustain higher rotation ($\propto V$) and random motions ($\propto \sigma$) than lower mass galaxies.}\label{fig:vsigprof}
\end{center}
\end{figure*}

\begin{figure*}
\begin{center}
\includegraphics[width=150mm]{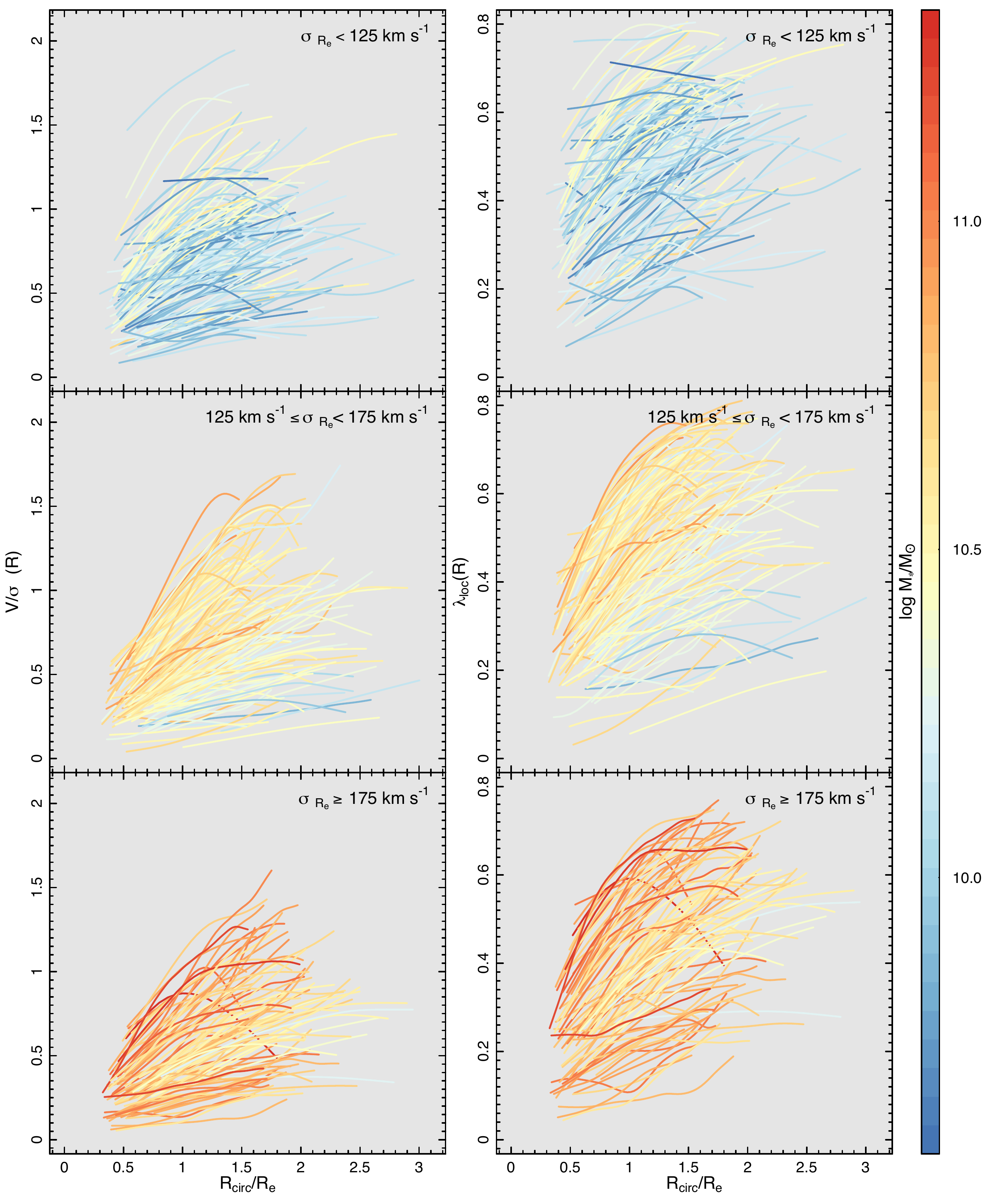}
\caption{Smoothed radial $V/\sigma(R)$ (left) and $\lambda_{\rm loc}(R)$ (right) profiles with measured stellar kinematics beyond $1.5R_e$. For clarity, the sample has been divided into three relatively equal-sized velocity dispersion bins: a low dispersion ($\sigma_{R_e}<125$ km s$^{-1}$, top), median dispersion ($125$ km s$^{-1}\le\sigma_{R_e}<175$ km s$^{-1}$, middle) and high dispersion ($\sigma_{R_e}>175$ km s$^{-1}$, bottom) bins as labelled. Profiles are colour-coded according to the galaxy stellar mass ($\log M_{*}$) as indicated by the colour scale on the far right. Most profiles show a steep inner rise with some reaching a plateau in the outskirts.}\label{fig:vosprof_mass}
\end{center}
\end{figure*}

In order to look for possible kinematic transitions in the outskirts of SAMI galaxies, we measure radial variations in the ratio of ordered to random stellar motion. For this purpose and to avoid weighting excessively against the fainter outskirts, we compute $V/\sigma(R)$ \emph{locally} following e.g. \citet{Wu14}:
\begin{equation}
V/\sigma(R)=\sqrt{\sum\limits_{i\in A} \left( \frac{V_i}{\sigma_i} \right) ^2}
\end{equation}
where $V_i$ and $\sigma_i$ are the measured recession velocity and velocity dispersion of the $i^{\rm th}$ spaxel in the elliptical annulus ($A$) with circularised radius $R$.

To ease comparison with literature, we also measure the \emph{local} spin parameter:
\begin{equation}
\lambda_{\rm loc}(R)=\frac{\sum\limits_{i\in A} F_i R_i |V_i|}{\sum\limits_{i\in A} F_i R_i \sqrt{V_i^2 + \sigma_i^2}}
\end{equation}
as adapted from \citet{Emsellem07} and defined by \citet{Bellstedt17a}. 

We choose rolling bin sizes of 25 spaxels at 10 spaxels intervals and remove 24 galaxies that have unreliably short profiles with fewer than 7 bins. This brings our final sample to 384 galaxies. The profiles are smoothed using a spline to minimise the effect of stochastic grid sampling.

The measured $V$, $\sigma$, $V/\sigma$ and $\lambda_{\rm loc}$ profiles are shown in Figs. \ref{fig:vsigprof} and \ref{fig:vosprof_mass}. Velocity and dispersion profiles show a clear trend with stellar mass in Fig. \ref{fig:vsigprof} such that more massive galaxies can support higher rotation and dispersion on average. Velocity profiles typically either rise monotonically, rise and fall or plateau with radius. As observed by \citet{FalconBarroso17} for cumulative dispersion profiles, we find a variety of local dispersion profiles. The vast majority ($\sim$96 percent) of dispersion profiles fall monotonically. We do observe a small fraction ($\sim$4 percent) of monotonically rising dispersion profiles or profiles that fall and rise. The most prominent cases are associated with galaxies with unusual kinematic features such as embedded (see Section \ref{sec:ED}) and counter rotating discs (also known as 2$\sigma$ galaxies). In Fig. \ref{fig:vosprof_mass}, we have divided the $V/\sigma$ profiles into three velocity dispersion bins chosen to optimise clarity. Most $V/\sigma$ and $\lambda_{\rm loc}$ profiles show a steep rise in the inner parts. Many profiles plateau at increasing radius. In all dispersion bins, we note a large range in $V/\sigma$ and $\lambda_{\rm loc}$ values at a given radius even among galaxies of similar mass. The middle row of Fig. \ref{fig:vosprof_mass} also shows that high mass galaxies on average have higher $(V/\sigma)$ and $\lambda_{\rm loc}$ profiles. This trend with stellar mass is especially noticeable in the intermediate velocity dispersion bin (i.e. middle panel of Fig. \ref{fig:vosprof_mass}).

\begin{figure*}
\begin{center}
\includegraphics[width=150mm]{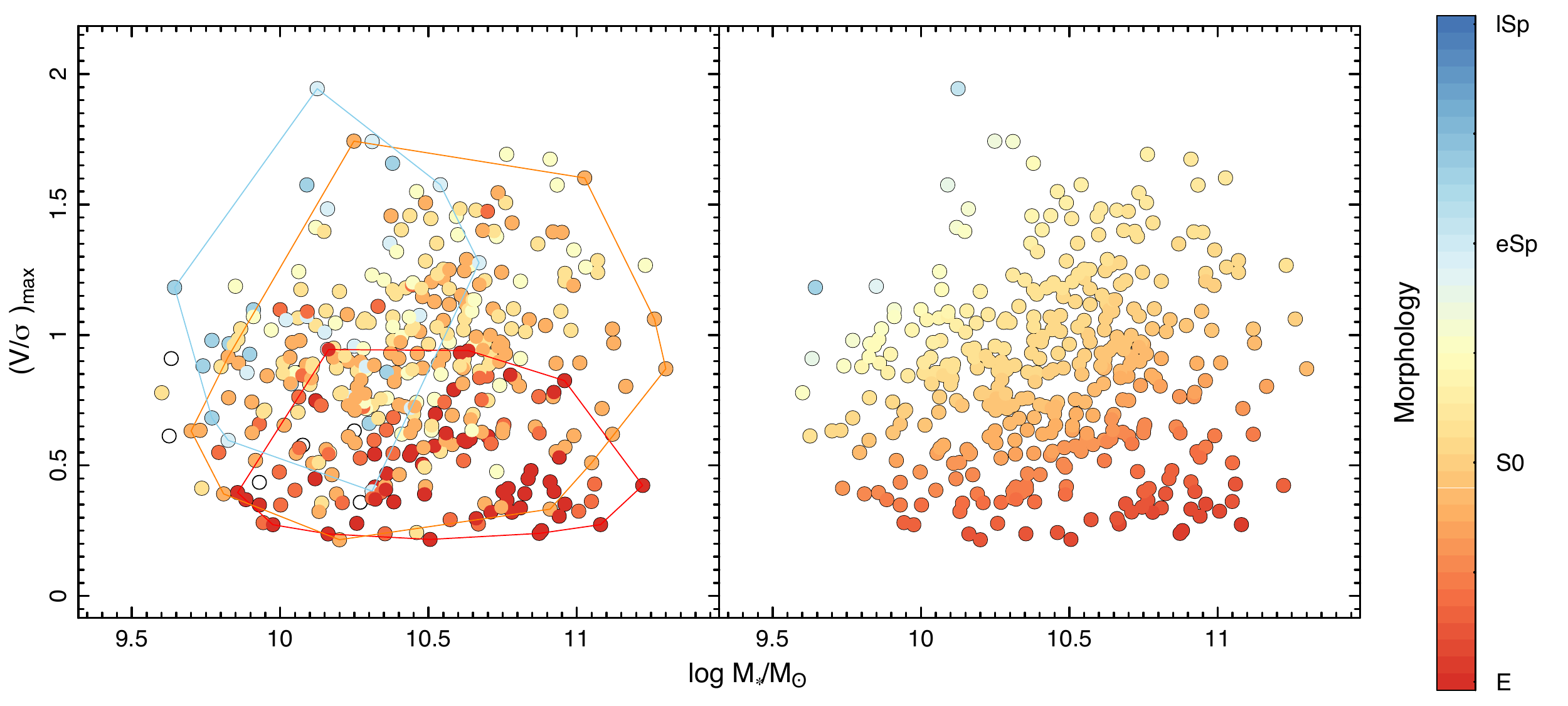}
\caption{Maximum measured $V/\sigma$ value, $(V/\sigma)_{\rm max}$ as a function of stellar mass (left). Data points are colour coded as a function of the galaxy visual morphology (E -- elliptical, S0 -- lenticular, eSp -- early spiral, lSp -- late spiral). Hollow symbols represent galaxies without a reliable morphological classification. Red, orange and blue polygons show the convex hull of E, S0 and Sp galaxies, respectively. The right hand panel shows the same distribution smoothed using a 2D polynomial {\sc loess} algorithm. Ignoring the colours, there is no overall trend with stellar mass, although a clear colour gradient can be seen in this space with significant overlap between the various morphological classes. 
}\label{fig:Mvos}
\end{center}
\end{figure*}

For the sake of simplicity, and to avoid unnecessary redundancy, we present results for $V/\sigma$ only henceforth. Our conclusions do not change qualitatively for $\lambda_{\rm loc}$. 

Fig. \ref{fig:Mvos} shows the maximum measured $V/\sigma$ value, $(V/\sigma)_{\rm max}$, as a function of stellar mass ($\log M_*/M_\odot$). While there is no clear correlation between $(V/\sigma)_{\rm max}$ and stellar mass (when ignoring the colours), the position of galaxies in this space is clearly correlated with visual morphology, albeit with a large range in $(V/\sigma)_{\rm max}$ at fixed morphology. Late-type galaxies (eSp and lSp) typically exhibit higher $(V/\sigma)_{\rm max}$ and tend to have lower masses than early-types (E and S0). 

\begin{figure}
\begin{center}
\includegraphics[width=90mm]{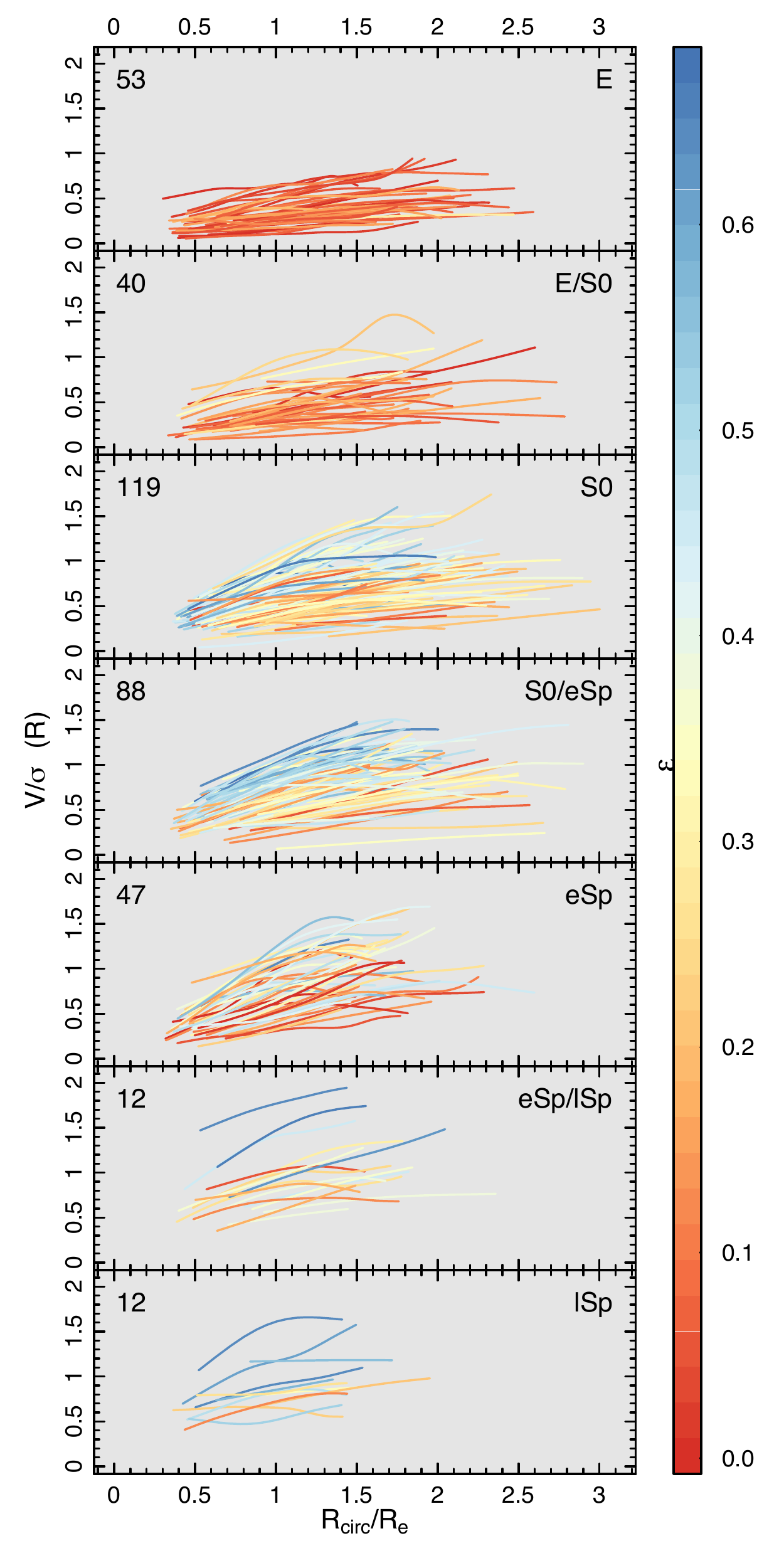}
\caption{Smoothed radial $V/\sigma(R)$ profiles with measured stellar kinematics beyond $1.5R_e$. The sample has been divided according to visual morphology \citep{Cortese16} from earlier (top) to later types (bottom) as labelled. Galaxies with no available visual morphologies are not shown. Profiles are colour coded according to their apparent ellipticity with roundest galaxies in blue and flatter galaxies in red. Number of galaxies is given in the top left for each panel. Maximum values of $V/\sigma(R)$ are lower for earlier types than for later types, indicating a lower level of rotational support in the former as expected.}\label{fig:vosprof_morph}
\end{center}
\end{figure}

The $V/\sigma$ profiles binned by visual morphology are shown in Fig. \ref{fig:vosprof_morph}. The amplitude of the $V/\sigma$ profiles is correlated with the apparent ellipticity of the system such that more apparently flattened systems have $V/\sigma$ profiles that lie above those of rounder ones. Since nearly all galaxies in SAMI are oblate axisymmetric systems \citep{Foster17}, the apparent ellipticity reflects a combination of the inclination of the system and its intrinsic flattening. Hence, the wide range in measured $V/\sigma$ values is partly explained by the fact that SAMI galaxies are viewed from a range of inclinations. Differences in seeing conditions is a further contributor to this scatter \citep[see e.g.][]{vandeSande17a,vandeSande17b}. Lenticular and late-type spiral galaxies overall show more rotational support than elliptical galaxies at all radii.

\begin{figure}
\begin{center}
\includegraphics[width=90mm]{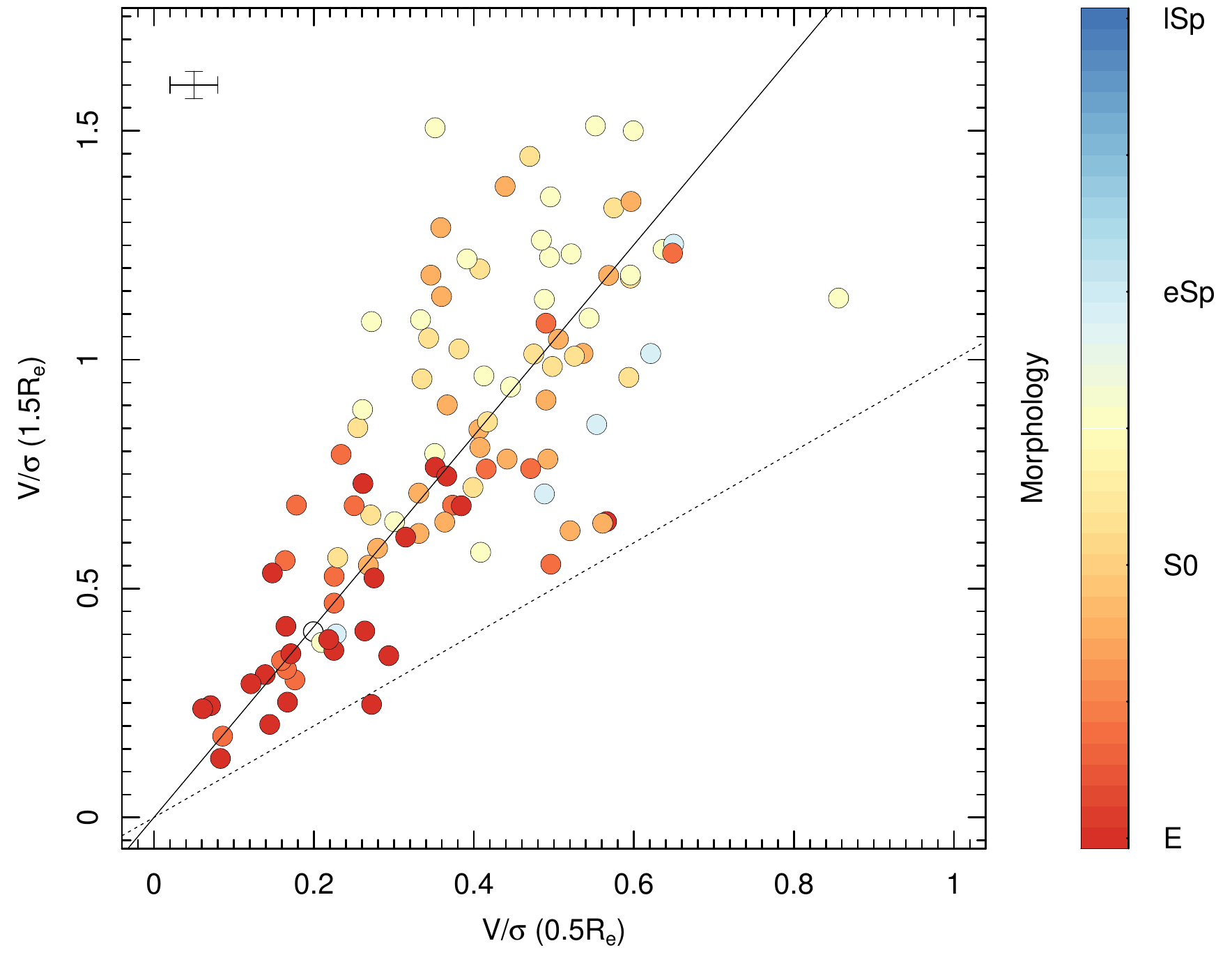}
\caption{Comparison of the inner ($0.5R_e$) and outer ($1.5R_e$) $V/\sigma(R)$. Points are colour coded by morphology. Hollow symbols represent galaxies without a reliable morphological classification. Solid and dashed lines represent a fit to the data and the one-to-one, respectively. There is a gradual departure from the one-to-one as $V/\sigma(R)$ increases suggesting that the $V/\sigma$ gradient is steeper in faster in systems that have high rotational support. The error bar represents the typical standard deviation about a smoothed profile.}\label{fig:vosgrad}
\end{center}
\end{figure}

\begin{figure}
\begin{center}
\includegraphics[width=90mm]{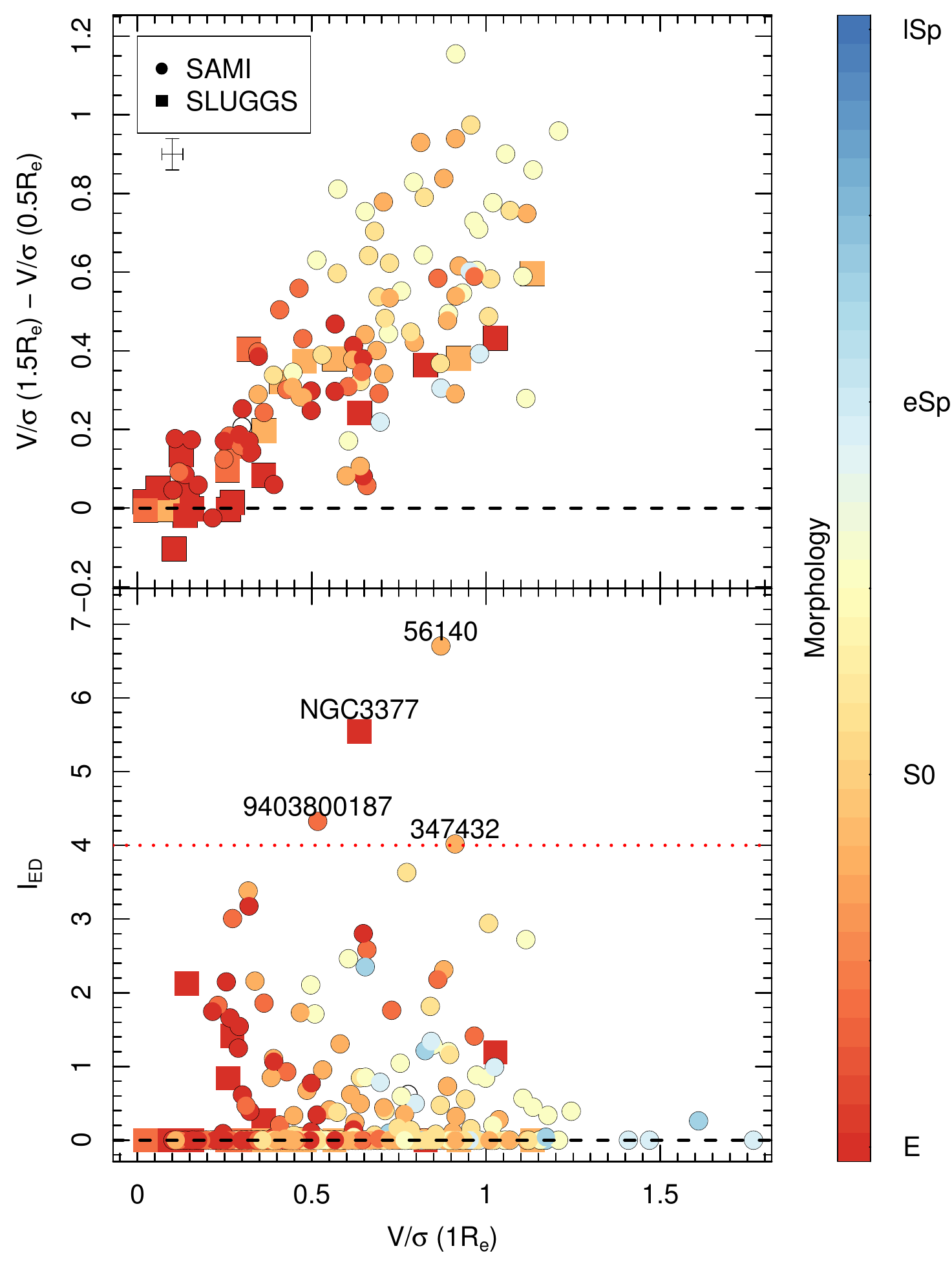}
\caption{Top panel: distribution of SAMI (circles) and ``SAMI-fied'' (with 1.5 arcsec seeing) SLUGGS (squares) galaxies in $V/\sigma(1R_e)$ vs $V/\sigma(1.5R_e)-V/\sigma(0.5R_e)$ space colour coded by morphology. Hollow symbols represent galaxies without a reliable morphological classification. On average, late-type galaxies have steeper gradients and higher rotational support than early-type galaxies. The error bar shown represents the typical standard deviation about the smoothed profiles. Bottom panel: embedded disc index $I_{\rm ED}$ (see Section \ref{sec:ED}) as a function of $V/\sigma(1R_e)$. Visually identified embedded discs (labelled) lie above the $I_{\rm ED}=4$ dotted red line.}\label{fig:vosgradvos}
\end{center}
\end{figure}

Fig. \ref{fig:vosgrad} compares the inner ($0.5R_e$) and outer ($1.5R_e$) $V/\sigma$ for SAMI galaxies with measured kinematic profiles at both radii (103 galaxies, refer to Fig. \ref{fig:vosprof_mass}). A similar trend was shown between 1 and $0.5R_e$ in \citet{vandeSande17b}. The data points lie slightly above the one-to-one line at low $V/\sigma$ and increasingly depart from it for higher $V/\sigma$ values. Galaxies with low inner $V/\sigma$ tend to have comparable $V/\sigma$ values in their outskirt, while galaxies with higher inner $V/\sigma$ show progressively steeper positive gradients. A linear regression fit to the data with intercept set to zero yields a slope of $2.09\pm0.06$, indicating significant departure from a slope of unity. There are very few outlier points in this space, suggesting that dramatic kinematic transitions between 0.5 and 1.5$R_e$ are rare in our sample. 
In Fig. \ref{fig:vosgradvos}, we perform an analysis akin to that presented in \citet[][their fig. 12]{Bellstedt17a} for $\lambda_{\rm loc}(R)$ and show the $V/\sigma$ gradient as a function of $V/\sigma(1R_e)$, colour coded by morphology.  We confirm a trend with morphology as reported in \citet{Bellstedt17a} that Es tend to have lower $V/\sigma$ values  and shallower $V/\sigma$ gradients than S0s. While our sample has few Sp galaxies, they tend to lie below the sequence defined by Es and S0s in Fig. \ref{fig:vosgradvos} (top panel). Apart from one notable exception, most galaxies are consistent with either a null or positive $V/\sigma$ gradient.

\begin{figure}
\begin{center}
\includegraphics[width=90mm]{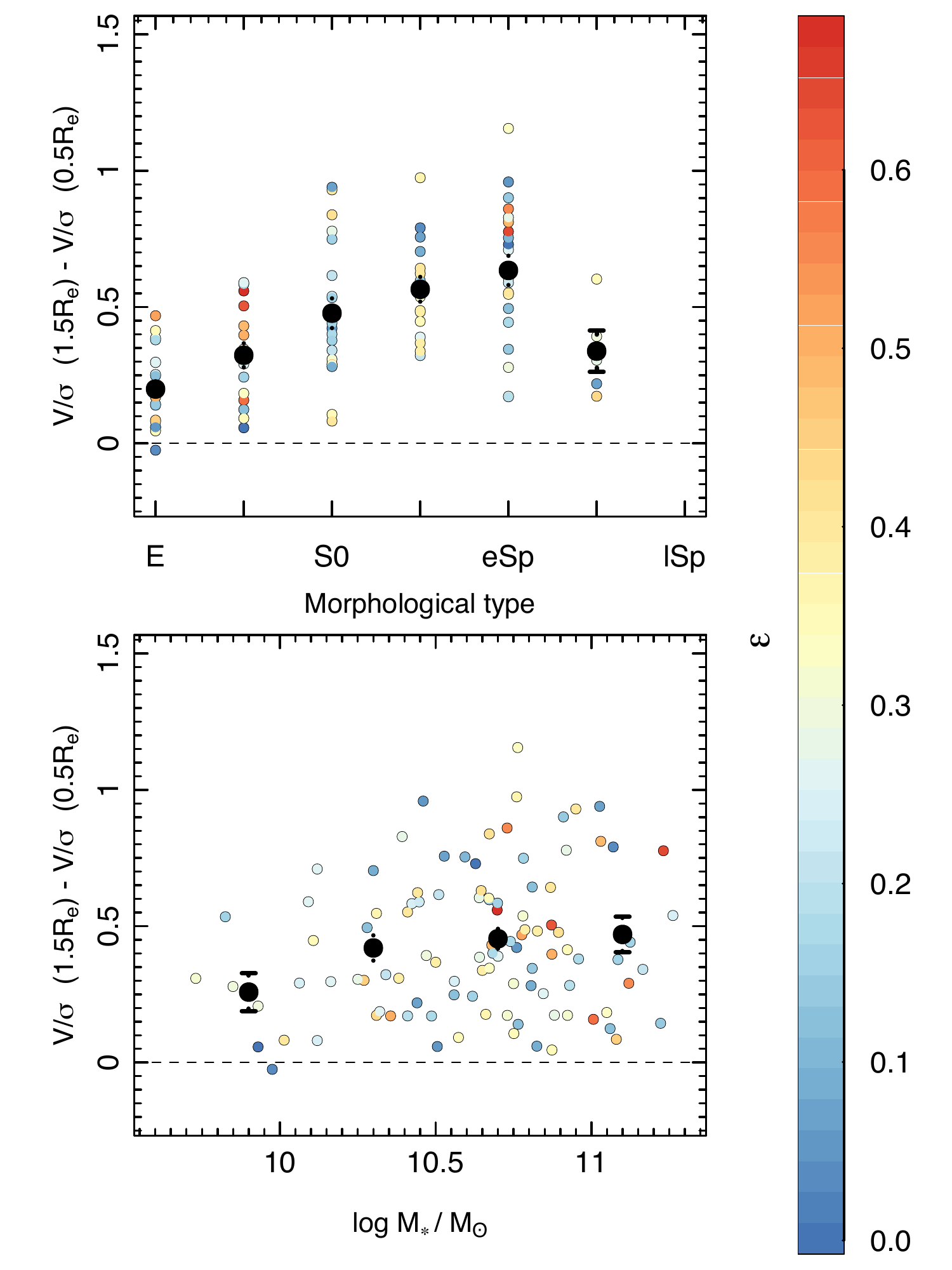}
\caption{$V/\sigma$ gradient versus morphology (top) and stellar mass (bottom) with gradient measured between $R=1.5R_e$ and $R=0.5R_e$. Small symbols represent individual galaxies and are coloured according to their apparent ellipticity ($\epsilon$). Galaxies without a reliable morphological classification are not shown in the top panel. Large black data points with error bars represent the average and standard error in each morphological class or stellar mass bin. There is a trend between the $V/\sigma$ gradient and morphology, although there is significant scatter within a given morphological class. The $V/\sigma$ gradient does not significantly correlate with stellar mass, although the lowest mass bin has a lower average $V/\sigma$ gradient. The lack of a correlation with ellipticity (i.e. colour gradient) suggests that projection effects are unlikely to explain the majority of the scatter.}\label{fig:vosgradmorph}
\end{center}
\end{figure}

In Fig. \ref{fig:vosgradmorph}, we show the $V/\sigma$ gradient between $1.5R_e$ and 0.5$R_e$ as a function of visual morphology and stellar mass. Despite the large scatter, there is a clear trend between the $V/\sigma$ gradient and morphology such that early-type galaxies have shallower gradients than late-types as found by \citet{Arnold11} and \citet{Foster16}. The lowest stellar mass bin has slightly lower average $V/\sigma$ gradient than the higher stellar mass bins.

\section{Discussion}\label{sec:discussion}

In this work, we explore the outer kinematic properties of galaxies in the SAMI Galaxy Survey. We specifically look at how well visual morphology maps onto the local rotation dominance parameter profiles. We also look for possible kinematic transitions between the in-situ and accreted regime of galaxy formation and embedded discs by identifying radial changes and modulations of the rotation dominance parameter in galaxies. 

\subsection{Large scale radial kinematic profiles and transitions}

Nearly all ($\gtrsim99$ percent) kinematic profiles shown in Figs. \ref{fig:vsigprof} and \ref{fig:vosprof_mass} are monotonic with many plateauing. Profiles exhibit a wide range in $V/\sigma(R)$, especially at large radii. 

Very few profiles exhibit the sudden kinematic transition expected between in-situ and accreted stars. If a result of 2-phase galaxy formation, this kinematic transition should be common and thus may be a distinct phenomenon to embedded discs, which will be discussed in Section \ref{sec:ED}. If a kinematic transition occurs, it may be very gradual such that in-situ and accreted stars are well mixed at intermediate radii or it may occur beyond the radii probed in this study ($>>1.5R_e$). \citet{Pulsoni17} recently observed a relationship between the transition radius and stellar mass. They detect the transition within $1R_e$ for $\log M_*/M_\odot\sim11.25$ and at progressively larger radii for lower mass galaxies, with galaxies of masses $\log M_*/M_\odot\sim10.5$ (the median radius in our sample) having an expected transition radius beyond 2.5$R_e$.

Our results confirm that the outer dynamical support reflects a galaxy's structure as identified through visual morphology (see Figs. \ref{fig:Mvos} and \ref{fig:vosprof_morph}). At fixed intrinsic flattening, $V/\sigma$ is expected to be larger and more readily measurable for edge-on systems than for face-on targets. Conversely, at fixed inclinations, we expect more intrinsically flattened systems to show higher rotations \citep{Foster17}. These trends are readily observable in the dependence of $V/\sigma (R)$ on apparent ellipticity shown in Fig. \ref{fig:vosprof_morph}, which is only detected in late-type systems. Because the inclination angle mostly affects the apparent dynamics of discy / intrinsically rotating systems, we observe that the range in $V/\sigma(R)$ measured in late-type systems is wider than among early-type systems at all radii.

\subsection{Outer balance in dynamical support}

Fig. \ref{fig:Mvos} shows significant overlap between Sp and S0 galaxies. However, the locus of Sp galaxies has higher rotational support and lower masses on average than the S0s. We here discuss the possible formation of lenticular galaxies through secular evolution of today's spiral galaxy population. Should one of today's spiral galaxies evolve in isolation, its velocity dispersion would increase slightly due to dynamical heating over time, while it might continue to form stars until it has exhausted its available gas reservoir \citep[also see][]{Cortesi13}. Hence, secular evolution of spiral galaxies would lead to lower $(V/\sigma)_{\rm max}$ and higher $\log (M_*/M_\odot)$, which would move the locus of spiral galaxies towards that of S0s in Fig. \ref{fig:Mvos}. However, the typical gas fraction (HI) around $\log M_*/M_\odot\sim 10$ is below unity \citep{Catinella12}, which would be insufficient to explain the most massive S0s in our sample. We note however that the progenitors of today's S0s need not look like today's spirals and may have indeed contained higher gas fractions \citep[e.g.][]{Popping14}.  Keeping this caveat in mind, the position of spiral and lenticular galaxies in $(V/\sigma)_{\rm max}$ -- $\log M_*/M_\odot$ space suggests that secular evolution alone may not be sufficient to explain the full range of properties observed in today's lenticular galaxies \citep[also see][]{FalconBarroso14,Querejeta15a}.

\citet{Fall83} showed that there is a correlation between the specific angular momentum $j*$ and the stellar mass of a galaxy. \citet{Romanowsky12}, \citet{Fall13} and \citet{Cortese16} also showed that the relationship for early-type galaxies is parallel but offset to lower $j*$ values or higher $\log (M_*/M_\odot)$ compared to that of late-type galaxies. 
Different from angular momentum, the rotation dominance parameter $(V/\sigma)$ is another measure of the prevalence of rotation in a galaxy. Fig. \ref{fig:Mvos} shows no overall correlation between stellar mass and $V/\sigma$. The lack of a correlation emphasises the fact that the rotation dominance parameter and angular momentum are not equivalent quantities \citep{Romanowsky12}. While there is no overall correlation between $(V/\sigma)_{\rm max}$ and stellar mass, there is a clear correspondence between the visual morphology of galaxies (symbol colours in Fig. \ref{fig:Mvos}) and their position in this space. 

The smoothed version of Fig. \ref{fig:Mvos} (right-hand panel) suggests that the slope of the relationship between $(V/\sigma)_{\rm max}$ and $\log (M_*/M_\odot)$ may vary smoothly with morphology. Given the potential sampling biases in this figure (e.g. no late type spiral galaxies), we do not attempt to fit a 3D surface to these data. Early(late)-types have shallower (steeper) relationships. We also note that despite a correlation between $R_{\rm max}$ and stellar mass, the trend is qualitatively the same when plotting $V/\sigma(0.5R_e)$ or $V/\sigma(1.5R_e)$ instead of $V/\sigma(R_{\rm max})$. The varying slopes in Fig. \ref{fig:Mvos} for various morphological types can be understood through the combination of the Faber-Jackson \citep[FJR,][]{Faber76} and the Tully-Fisher \citep[TFR,][]{Tully77} Relations, which relate the rotational velocity and the velocity dispersion to the stellar mass in discs and bulges, respectively. 
The TFR suggest that in disc dominated galaxies (i.e. $\sigma < V_{\rm rot} \propto V$, blue points), higher mass galaxies should have higher rotational velocities, hence the numerator of the y-axis is higher for higher masses. On the other hand, for bulge dominated galaxies (i.e. $\sigma > V_{\rm rot}$, red points), the FJR suggests that velocity dispersion correlates with stellar mass, hence the higher the mass, the higher the denominator on the y-axis, yielding lower $(V/\sigma)_{\rm max}$. So we expect a positive relation for disc dominated galaxies and a negative (rational) relation for bulge dominated galaxies (i.e. shallow and/or negative slope) as observed. Intermediate morphologies have intermediate positions in this space. 

Inclination is likely to contribute significantly to the mixing of the morphological classes in Fig. \ref{fig:Mvos} by scattering points down in that space. Despite the confusion created by random inclinations, the dependence on morphology is still clearly seen in this space, so inclination is not enough to completely erase this trend. Correcting for inclination would require an assumption on the intrinsic shape of individual galaxies that also correlates with morphology, thereby adding significant systematic uncertainties to the trend.

\subsection{Search for embedded discs}\label{sec:ED}

\begin{figure*}
\begin{center}
\includegraphics[width=180mm]{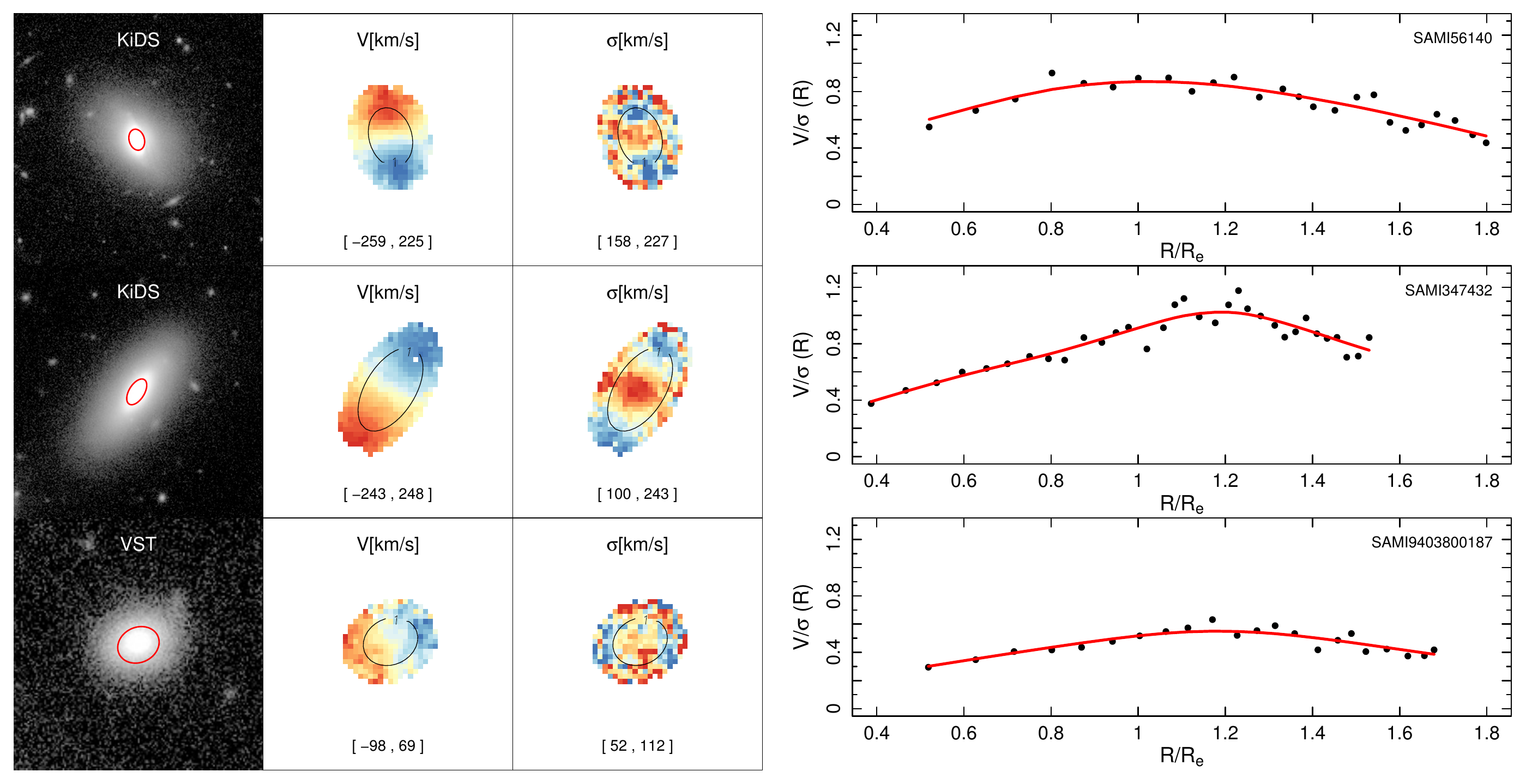}
\caption{From left to right, $r$-band image, velocity and velocity dispersion maps, and $V/\sigma$ profile for galaxies with visually identified embedded discs in our sample. Red and black ellipses represent 1$R_e$. All three galaxies show rising and declining $V/\sigma$ profile in the inner and outer parts, respectively. The inner parts of both SAMI56140 and SAMI347432 appear discy on the images. The outer spectra for SAMI9403800187 show significant contamination from a nearby star (see text). The plotted range of values [blue, red] is given at the bottom of each kinematic map. For each row, the galaxy ID is labelled in the top right corner. Red lines show the profiles smoothed over numerical artefacts in the right-hand panel.}\label{fig:ED}
\end{center}
\end{figure*}

\begin{figure*}
\begin{center}
\includegraphics[width=180mm]{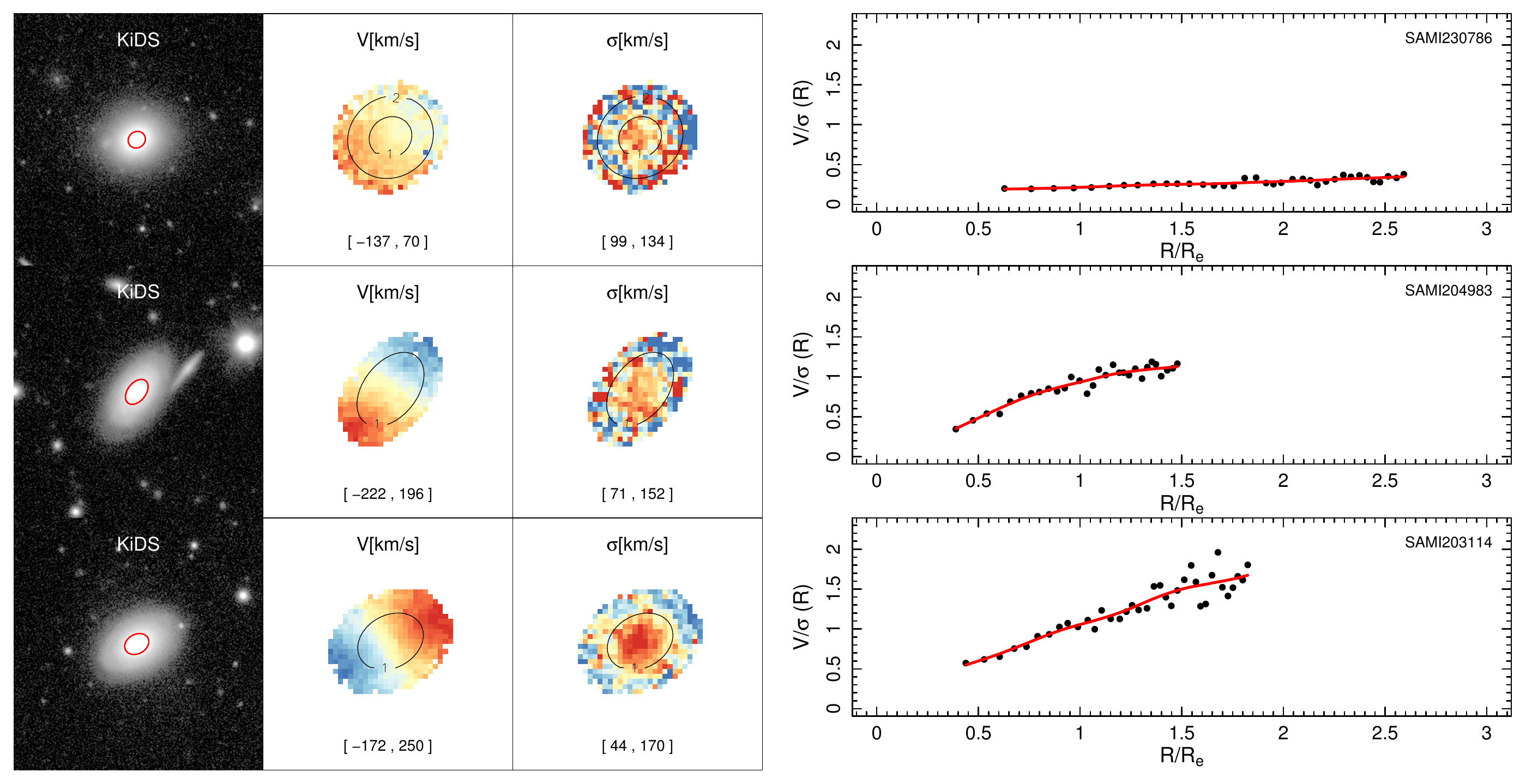}
\caption{From left to right, $r$-band image, velocity and velocity dispersion maps, and $V/\sigma$ profile for three typical galaxies with a range of $V/\sigma$ amplitudes. Ellipses represent integer values of $R_e$ as labelled. The plotted range of values [blue, red] is given at the bottom of each kinematic map. For each row, the galaxy ID is labelled in the top right corner. Red lines show the profiles smoothed over numerical artefacts in the right-hand panel.}\label{fig:TG}
\end{center}
\end{figure*}

\citet{Bellstedt17a} found significantly negative $\lambda_{\rm loc}$ gradients in 4 of their 28 early-type galaxies. These outer negative $\lambda_{\rm loc}$ gradients often signal the presence of a discy structure embedded within a larger-scale bulge, so-called ``embedded discs''. \citet{Arnold14} visually identified 6/22 galaxies with embedded discs in the SLUGGS survey. Using the beta distribution to estimate the 68 percent confidence interval as recommended by \citet{Cameron11}, the latter implies that (20--38) percent of galaxies in SLUGGS harbour embedded discs.  
These embedded discs have recently been related to ellicular galaxies \citep[ES,][]{Liller66,Graham16}, which are galaxies of intermediate E/S0 morphologies. 

We perform a visual inspection using diagnostic plots similar to those shown in Fig. \ref{fig:ED} and \ref{fig:TG} to identify galaxies with convincingly declining outer $V/\sigma(R)$ profiles. Identified embedded discs must have visibly declining smoothed profiles when considering the scatter of individual bins. This visual inspection of individual profiles and kinematic maps reveals that there are only 3/384 SAMI galaxies with convincingly declining outer $V/\sigma(R)$ profiles in our sample (see Fig. \ref{fig:ED}): SAMI56140 (S0), SAMI347432 (S0) and SAMI9403800187 (E/S0). These numbers correspond to a 68 percent confidence interval of (0.5--1.5) percent for SAMI galaxies with embedded discs, which is at least an order of magnitude fewer than in the SLUGGS sample. In Fig. \ref{fig:ED}, SAMI56140 and SAMI347432 both show lower velocity dispersion along the major than the minor axis and discy features on the images, which are independent signatures of embedded discs \citep{Arnold14}. We note that spectra for SAMI9403800187 are contaminated by scattered light from a nearby bright star of A spectral type. Explicitly fitting for this contamination reveals that it is significant and has artificially enhanced the velocity dispersion beyond $1R_e$, making this a less secure embedded disc detection.

The SLUGGS survey is comprised of 25+3 carefully selected early-type galaxies chosen to be representative of the underlying distribution of galaxy properties \citep{Brodie14}. For the purposes of the globular clusters studies, edge-on inclinations were favoured, which may have had an impact on the potentially higher detectability of embedded discs. In contrast, SAMI makes no explicit selection on morphology or inclination. 

In order to quantitatively explain the tension between the frequency of embedded discs found in SAMI and SLUGGS, we mimic the observational imprint of SAMI onto the SLUGGS stellar kinematic maps from \citet{Foster16}. This ``SAMI-fication'' involves resampling of the velocity and dispersion maps to the same pixel scale as SAMI, assuming a distance 20 times further away to mimic the differences in both spatial sampling and median distances of the two surveys. Finally, we convolve the corresponding maps with a 1.5 and 2.5 arcsec Gaussian smoothing kernel, \emph{propagating and re-fitting the line-of-sight velocity distribution in each spaxel} to mimic good and typical SAMI seeing, respectively. The smoothing leads to unrealistically low spaxel-to-spaxel scatter and we do not attempt to explicitly model complex measurement uncertainties. Of the 6 SLUGGS embedded discs identified in \citet{Arnold14}, none and one (NGC~3377) galaxy has a visibly declining $V/\sigma(R)$ profiles in 2.5 and 1.5 arcsec seeing, respectively, within the SAMI field-of-view. The $V/\sigma(R)$ profile of NGC~3377 peaks at a radius around 1.8$R_e$ for the 1.5 arcsec seeing case. With a stellar mass of $\log M_*/M_\odot=10.50$ \citep{Forbes17}, this galaxy would have made the SAMI selection cut at a median redshift of $\sim0.06$ \citep{Bryant15}. 
Only 66 percent of SAMI galaxies within $10.5<\log M_*/M_\odot<11$ have measured stellar kinematics beyond 1.8$R_e$, and accounting for the fraction of early-types in our sample (80 percent) this corresponds to an expected fraction of (0--11) percent of galaxies harbouring detectable embedded discs. In summary, after accounting for differences in the sample selection, observational and instrumental effects as described above, the embedded disc fractions of SAMI and SLUGGS are consistent.


To address the subjectivity associated with the visual inspection of the kinematic profiles in identifying embedded discs, we propose the following ``embedded disc index'':
\begin{equation}
I_{\rm ED}\equiv\frac{(V/\sigma)_{\rm max} - V/\sigma(R_{\rm max})}{(V/\sigma)_{\rm sd}},
\end{equation}
where $V/\sigma(R_{\rm max})$ is the outermost value and $(V/\sigma)_{\rm sd}$ is the standard deviation about the smoothed profile. As seen in Fig. \ref{fig:vosgradvos}, this clearly segregates galaxies that harbour visually identified embedded discs by highlighting the significance of declining outer $V/\sigma$ profiles. The visual threshold seems to be around $I_{\rm ED}=4$, i.e. an embedded disc is identified by eye if the profile declines by at least 4 times the scatter. Bootstrapping the kinematic maps 10,000 times, we find that the significance level of the declining profiles are $>99.99$, 99.3 and 99.1 for SAMI56140, SAMI347432 and SAMI9403800187, respectively. The latter number is likely overestimated due to the aforementioned scattered light contamination. Statistically significant $V/\sigma$ drops ($I_{\rm ED}\ge4$) combined with azimuthally varying dispersion maps feature are not seen in more typical galaxies (see Figs. \ref{fig:vosprof_mass} and \ref{fig:TG}). We note however that the visually identified embedded discs are simply the ``tip of the iceberg'' as there is a broad distribution of $I_{\rm ED}$.


\subsection{$V/\sigma$ gradients, morphology and mass}

The clear colour gradient seen in Fig. \ref{fig:vosgradvos} suggests that $V/\sigma$ gradients may be correlated with morphology. A similar trend was identified in the SLUGGS survey by \citet{Arnold11}, \citet{Foster16} and \citet{Bellstedt17a}, but was not seen in \citet{Raskutti14}. Fig. \ref{fig:vosgradmorph} shows that such a relationship exists in SAMI, but the scatter in individual $V/\sigma$ gradient measurements is remarkably large. The large scatter explains why small samples such as the SLUGGS survey and that of \citet{Raskutti14} came to opposite conclusions. One requires a survey of several hundreds of galaxies to be able to reliably identify this trend. We further note that the scatter in $V/\sigma$ gradients is not correlated with apparent ellipticity, suggesting that this scatter is not principally caused by random inclination angles. The large scatter could be a mixture of measurement uncertainties and intrinsic differences between individual galaxies.

Since bulges have higher pressure support than discs and discs have higher rotational support than bulges, this trend could reflect a gradual change in the transition radius between the pressure and rotation supported regime \citep[e.g.][]{MendezAbreu18}. Ellipticals tend to have prominent bulges and shallow gradients because there is no outer disc to transition to. On the other hand, galaxies with two components (S0 and eSp) have comparatively much steeper gradients, suggesting a transition from pressure to rotational support occurs between 1.5$R_e$ and $0.5R_e$. Fig. \ref{fig:vosgradmorph} suggests that this trend may flatten out or turn over in late-type spirals (lSp) such that lSp galaxies have either slightly shallower or similar gradients than S0 and eSp galaxies. This behaviour may be related to the observed constant velocity dispersion profiles found in pseudobulges by \citet{Neumann17}. It may also be related to the tendency for later type spirals to have overall lower spin parameters than their earlier-type counterparts noticed by e.g. \citet{FalconBarroso14}. 
Constant $V/\sigma$ profiles amongst late-type spirals may also indicate that the bulge typically is confined to radii $\lesssim0.5R_e$, and hence the pressure supported regime is not well probed in this Fig. \ref{fig:vosgradmorph}.

As was noted earlier, Figs. \ref{fig:vosprof_mass} and \ref{fig:vosgradmorph} suggest that the amplitude and gradient of the $V/\sigma$ profiles depend on the stellar mass of galaxies. More massive systems tend to have both steeper and higher $V/\sigma$ profiles. Fig. \ref{fig:vsigprof} shows that the rotational support increases and the pressure support decreases with stellar mass on average in our sample. The velocity profiles usually have a positive slope that tends to be steeper for higher mass systems. Conversely, dispersion profiles are usually negative with more negative slopes in more massive systems. Hence, the weak trend seen in Fig. \ref{fig:vosgradmorph} is a combination of both steeper positive velocity and negative dispersion profiles.

\section{Conclusions}\label{sec:conclusion}

In this work, we study the balance of rotational to pressure support in the outskirts of 384 SAMI galaxies with stellar kinematics extending to $\ge1.5R_e$. While we mainly show $V/\sigma(R)$ profiles, our results do not change qualitatively for $\lambda_{\rm loc}$. Our conclusions are summarised below.

\begin{enumerate}

\item We confirm that the structure of galaxies, as probed through their visual morphology, is linked to their dynamical support and we find this to be true at all probed radii. On average, visually discy systems have higher rotational support and steeper $V/\sigma$ gradients than bulge dominated systems. In other words, the visual morphology of galaxies reflects their dynamical state, albeit with large intrinsic scatter and some possible confusion due to inclination. 

\item We identify three galaxies out of a sample of 384 with convincing embedded discs in the SAMI sample. The fraction of embedded discs in SAMI is roughly an order of magnitude below that of SLUGGS \citep{Arnold11,Bellstedt17a}. Once accounting for differences in the distances, spatial sampling and observational effects, the fractions of embedded discs in both surveys are brought into agreement.

\end{enumerate}

The SAMI Galaxy Survey stellar kinematic sample will double in size this year as the recently observed data are analysed and added to the existing sample. In a follow-up paper (Foster et al. in prep), we explore radial tracks in the spin-ellipticity diagram  \citep{Graham16} to further investigate the link between visual structure and kinematics. The planned multi-object wide-field IFU spectrograph, HECTOR \citep{Bryant16} on the Anglo-Australian Telescope is designed to have optimised radial coverage of the stellar kinematics out to at least $2R_e$. The associated ambitious galaxy survey is better optimised to probe kinematic transitions and embedded disc fractions.

\section*{Acknowledgments}
This research was conducted by the Australian Research Council Centre of Excellence for All Sky Astrophysics in 3 Dimensions (ASTRO 3D), through project number CE170100013. The SAMI Galaxy Survey is based on observations made at the Anglo-Australian Telescope. The Sydney-AAO Multi-object Integral field spectrograph (SAMI) was developed jointly by the University of Sydney and the Australian Astronomical Observatory. The SAMI input catalogue is based on data taken from the Sloan Digital Sky Survey, the GAMA Survey and the VST ATLAS Survey. The SAMI Galaxy Survey website is http://sami-survey.org/. 

NS acknowledges support of a University of Sydney Postdoctoral Research Fellowship. JvdS is funded under Bland-Hawthorn's ARC Laureate Fellowship (FL140100278). M.S.O. acknowledges the funding support from the Australian Research Council through a Future Fellowship (FT140100255). SB acknowledges the funding support from the Australian Research Council through a Future Fellowship (FT140101166). Support for AMM is provided by NASA through Hubble Fellowship grant \#HST-HF2-51377 awarded by the Space Telescope Science Institute, which is operated by the Association of Universities for Research in Astronomy, Inc., for NASA, under contract NAS5-26555.

\bibliographystyle{mnras}
\bibliography{biblio}

\label{lastpage}

\end{document}